\documentstyle[aps,preprint,prb]{revtex}

\begin{document}
 
\title{A Numerical Renormalization Group Study of 
a Kondo Hole in a One Dimensional Kondo Insulator}
\author {Clare C. Yu\\
Department of Physics and Astronomy\\
University of California, Irvine\\
Irvine, CA 92717-4575}
\date{ \today}
 
\maketitle
\setcounter{page}{0}
\thispagestyle{empty}
\begin{abstract}
We have studied a Kondo hole in a
one-dimensional Kondo insulator at half-filling using a 
density matrix formulation of the numerical 
renormalization group. The Kondo hole introduces
midgap states. The spin density introduced by the hole
is localized in the vicinity of the hole. It
resides primarily in the f-spins for small exchange
coupling $J$ and in the conduction spins for
large $J$. We present results on the spin gap, charge gap,
and neutral gap. For small $J$,
the spin gap is smaller than the charge gap, while for
large $J$, the spin gap is larger than the charge gap.
The presence of the Kondo hole reduces RKKY interactions
as can be seen 
in the staggered susceptibility.
\end{abstract}
\vspace{1.0in}

\newpage
\thispagestyle{empty}
\section{Introduction}
The onset of coherence in a Kondo lattice is associated with
the opening of a gap in a Kondo insulator. Impurities can disrupt
this coherence and produce midgap states.
In this paper we will examine the effect of putting a Kondo hole in
a one-dimensional Kondo insulator. A Kondo hole is a nonmagnetic
impurity which has a conduction orbital but no f--orbital. 
Experimentally this is done by replacing 
Ce or U ions with La or Th ions. There is experimental evidence that
these nonmagnetic impurities can behave as Kondo impurities. 
For example,
CePd$_{3}$ is a good metal whose resistivity decreases with 
decreasing
temperature as $T$ approaches zero. However, when nonmagnetic La ions
are substituted for Ce ions 
in Ce$_{1-x}$La$_{x}$Pd$_{3}$, the resistivity below 50 K 
increases with decreasing temperature in a fashion 
reminiscent of Kondo impurities in a metal.\cite{jon}

Sollie and Schlottmann\cite{schlottmann} have done calculations on
a Kondo hole in an Anderson lattice
with the energy of the f--orbital 
$\varepsilon_{f}=\infty$ on the Kondo 
hole site and a finite value of $\varepsilon_{f}$ on the other sites.
They calculate the self--energy to second order perturbation
in $U$ about the Hartree--Fock solution, though they do not
calculate the self--energy self--consistently and they
neglect its momentum dependence. By
examining the local f--electron density of states, 
they find midgap states in the vicinity of the hole site.
However, this approach does not distinguish between the
various types of gaps, e.g., spin and charge.

In this paper we use the density matrix renormalization group 
approach
\cite{White} to study a Kondo hole in a one dimensional Kondo 
lattice.
To the best of our knowledge, this is the first numerical
calculation of a Kondo hole in a Kondo lattice.
The paper is organized as follows. In section II we present the
Hamiltonian, which we study using the density matrix renormalization
group approach.\cite{White} We present our results in section III.
In section IIIa, we discuss the chemical potential as a function
of electron filling. We find that the impurity introduces midgap
states which lie in the center of the quasiparticle gap for
large values of the exchange coupling $J$ and 
move towards the edges of the gap as $J$ decreases. In section IIIb,
we present our results on the spin gap, charge gap, 
and neutral gap as function of $J$. For small $J$ 
($J\stackrel{\textstyle <}{\sim}3t$),
the spin gap is smaller than the charge gap, while for
large $J$ ($J\stackrel{\textstyle >}{\sim}3t$), the spin 
gap is larger than the charge gap. In section IIIc we find that
the spin of the Kondo hole resides primarily in the conduction
spins for large $J$ and primarily in the f--spins for small $J$.
In section IIId we discuss the effect of the Kondo hole on the RKKY
interactions. By examining 
the staggered susceptibility, we find that the RKKY oscillations
are reduced when compared to a lattice with no Kondo hole.
We state our conclusions in section IV.

\section{Hamiltonian}
The standard one dimensional Kondo lattice has
spin-1/2 conduction electrons that hop from site
to site with an on--site spin exchange $J(i)$ between the f--electron
and the conduction electron on that site. In the midst of this
chain we place a Kondo hole which has no f--orbital, and hence,
no on--site exchange. Thus the Hamiltonian is
\begin{equation}
H=-t\sum_{i\sigma}\left(c^{\dagger}_{i\sigma}c_{i+1\sigma}
+h.c.\right)
+\sum_{i}J(i){\vec S}_{if}\cdot{\vec S}_{ic}
\end{equation}
where the conduction electron spin density on site $i$ is 
${\vec S}_{ic}=\sum_{\alpha\beta}c^{\dagger}_{i\alpha}
({\vec \sigma}/2)_{\alpha \beta}c_{i\beta}$, 
and ${\vec \sigma}_{\alpha\beta}$ are Pauli matrices.
On the host lattice the f--electron spin density is
${\vec S}_{if}=\sum_{\alpha\beta}f^{\dagger}_{i\alpha}
({\vec \sigma}/2)_{\alpha
\beta}f_{i\beta}$, while on the Kondo hole site ${\vec S}_{if}=0$ 
because
there is no f--electron. $t$ is the hopping matrix element 
for the conduction electrons between neighboring sites. We set
$t=1$.
The on--site spin exchange $J(i)$ is zero for the Kondo hole
and equal to $J$ on the rest of the sites. We choose
$J$ to be antiferromagnetic ($J>$0). We place the Kondo hole
in the middle of the lattice on site $i=L/2$, where L is the number
of sites. We studied lattices of size $L$=4, 6, 8, 16, and 24.
In the absence of a Kondo hole, the Kondo insulator corresponds to
half filling where the total number of conduction electrons $N$ 
equals the number of sites $L$.

Even with a Kondo hole, the Hamiltonian has SU(2) spin symmetry as 
well as SU(2) charge pseudospin symmetry.\cite{nish} 
The components of the pseudospin operator $\vec{I}$ are given by:
\begin{eqnarray}
\nonumber  I_{z} & = & \frac{1}{2} \sum_{i} 
(c^{\dagger}_{i \uparrow} c_{i \uparrow}+
            c^{\dagger}_{i \downarrow}c_{i \downarrow}+
            f^{\dagger}_{i \uparrow}f_{i \uparrow}+
            f^{\dagger}_{i \downarrow}f_{i \downarrow} -2) \\
            I_{+} & = & \sum_{i} (-1)^{i} (c^{\dagger}_{i \uparrow}
                                       c^{\dagger}_{i \downarrow}-
             f^{\dagger}_{i \uparrow}f^{\dagger}_{i \downarrow}) \\
\nonumber  I_{-} & = & \sum_{i} (-1)^{i} 
(c_{i \downarrow}c_{i \uparrow}-
        f_{i \downarrow}f_{i \uparrow})
\end{eqnarray}
The z--component of the pseudospin is the charge operator
and is equal to $(N_{el}/2) - L$, where $N_{el}$ is the 
total number of
electrons including both conduction and f--electrons.
An $I_{z}=1$ state can be achieved by adding two electrons.

All the energy eigenstates have a definite value of S and I.
At half--filling with one Kondo hole ($N=L$ and $N_{el}=2L-1$),
the ground state is a pseudospin singlet with total spin $S=1/2$
($S=1/2$, $I=0$) for all values of $J$. The spin gap $\Delta_{S}$
is defined as the energy difference between the lowest--lying
excited spin state and the ground state:
\begin{equation}
\Delta_{S}=E(S=\frac{3}{2},I=0)-E_{0}(S=\frac{1}{2},I=0)
\end{equation}
where $E_{0}$ is the energy of the ground state. For $J\gg t$,
the lowest spin excitation corresponds to forming a triplet 
between an f--spin and a conduction spin on a site that is
not a Kondo hole, with the remaining sites being the
same as in the ground state. In this limit $\Delta_{S}\cong J$.

To find the charge gap, we note that optical experiments 
measure the charge gap by measuring the conductivity 
which is determined by the current-current correlation 
function. The current is related to the charge density 
through the continuity equation. Thus the lowest
lying charge excitation is the lowest excited 
state $|n>$ with $S=1/2$ such that 
$<0|\sum_{q} \rho_{q}|n>\neq 0$, where $\rho_{q}$ 
is the q-component of the Fourier transformed charge 
density operator and $|0>$ is the ground 
state.\cite{guerrero} Notice that $\rho_{q}$ is related to 
$\vec{I}^{z}_{q}$, where $\vec{I}_{q}$ is a Fourier 
transformed vector in pseudospin space given by
\begin{eqnarray}
\nonumber  I^{z}_{q} & = & \frac{1}{2} \sum_{i} e^{-i\vec{q}
\cdot \vec{r}_{i}}
            (c^{\dagger}_{i \uparrow}c_{i \uparrow}+
            c^{\dagger}_{i \downarrow}c_{i \downarrow}+
            f^{\dagger}_{i \uparrow}f_{i \uparrow}+
            f^{\dagger}_{i \downarrow}f_{i \downarrow} -2) \\
            I^{+}_{q} & = & \sum_{i} e^{-i\vec{q}\cdot \vec{r}_{i}} 
                        (-1)^{i} (c^{\dagger}_{i \uparrow}
                                c^{\dagger}_{i \downarrow}-
            f^{\dagger}_{i \uparrow}f^{\dagger}_{i \downarrow}) \\
\nonumber  I^{-}_{q} & = & \sum_{i} e^{-i\vec{q}\cdot \vec{r}_{i}}
        (-1)^{i} (c_{i \downarrow}c_{i \uparrow}-
        f_{i \downarrow}f_{i \uparrow})
\end{eqnarray}
Using the Wigner-Eckart theorem, one can show that
the ($S=1/2$, $I=1$) states are the only states
$|n>$ for which the charge density $\rho_{q}$
has finite matrix elements $<n|\rho_{q}|0>$ with the
ground state $|0>$. Thus
the charge gap $\Delta_{C}$ is the energy
difference between the ground state and the lowest pseudospin
triplet state:\cite{guerrero,nish}
\begin{equation}
\Delta_{C}=E(S=\frac{1}{2},I=1)-E_{0}(S=\frac{1}{2},I=0)
\end{equation}

We also define a neutral 
gap $\Delta_{N}$ as the energy difference between the
ground state and the lowest--lying excited neutral
state with the same quantum numbers as the ground
state ($S=1/2$, $I=0$):\cite{ccysrw}
\begin{equation}
\Delta_{N}=E(S=\frac{1}{2},I=0)-E_{0}(S=\frac{1}{2},I=0)
\end{equation}
For the half--filled Kondo lattice without a Kondo hole,
the neutral singlet has been found to be an elementary
excitation consisting of a ``particle'' and a ``hole'', which are
($S=1/2$, $I=1/2$) excitations. In a single site basis,
a ``hole'' is a site with one f--electron and no conduction
electrons with quantum numbers ($S=1/2$, $I=1/2$, $I_{z}=-1/2$);
this hole is different from a Kondo hole. A ``particle'' is a site
with one f--electron and two conduction electrons with
($S=1/2$, $I=1/2$, $I_{z}=+1/2$). A particle and a hole
can be combined to form a charge excitation ($S=0$, $I=1$)
or a neutral singlet excitation ($S=0$, $I=0$).
(Other combinations are also possible.) When a Kondo hole
is added to the lattice, one can think in terms of a hole
or a particle on the Kondo hole site. A hole on the Kondo
hole site has no conduction electrons, no f--electrons,
and quantum
numbers ($S=0$, $I=1/2$, $I_{z}=-1/2$). A particle on the
Kondo hole site has 2 conduction electrons, no f--electrons,
and quantum numbers ($S=0$, $I=1/2$, $I_{z}=+1/2$). A particle
(hole) on a Kondo hole site can be combined with a hole (particle)
on an ordinary Kondo site to form a charge excitation 
($S=1/2$, $I=1$) or a neutral excitation 
($S=1/2$, $I=0$).\cite{neutral}
These are the excitations associated with the charge gap
and the neutral gap.
 
We use the density matrix
renormalization group (DMRG) algorithm \cite{White}
to calculate the ground state and the first few
excited states of the Kondo lattice. This real-space
technique has proven to be remarkably accurate for
one dimensional systems
such as the Kondo and Anderson lattices.\cite{guerrero,metal}
We used the finite system method \cite{White} with open
boundary conditions in which there is no hopping past the
ends of the chain. 
We kept up to 140 states per block. The results were extremely
accurate for $J\gg t$, with typical truncation 
errors of order $10^{-10}$
for $J$=10. For $J\stackrel{\textstyle <}{\sim}t$, the f-spin
degrees of freedom lead to a large number of nearly degenerate
energy levels. As a result, the
accuracy was significantly reduced, with truncation errors of
order $10^{-4}$ for $J$=0.5. 

\section{Results}
\subsection{Chemical Potential Versus Filling}
We study how the chemical potential varies with electron filling.
We consider a 16 site Kondo lattice with the Kondo hole on site 8.
We vary the electron filling and define the chemical potential by
\begin{equation}
\mu(N)=E_{0}(N)-E_{0}(N-1)
\end{equation}
where $E_{0}(N)$ is the ground state energy with $N$ electrons. Our 
results are shown in Fig. \ref{fig:chempot}, where we have scaled
the chemical potential by $J$. 
When the Kondo hole is absent,
there is a jump in the chemical potential that is centered about
half--filling ($N=16$). This is the quasiparticle gap which
is defined  as the difference of chemical potentials
\begin{equation}
\Delta_{QP}=\mu(N+1)-\mu(N)
\end{equation}
From Fig.
\ref{fig:chempot}, we see that the Kondo hole introduces
states in the gap
for large $J$. The chemical potential
of these midgap states corresponds to the energy of adding
a particle or a hole to the half--filled system.
To understand why these midgap states have a chemical potential
so close to zero, note that for $J\gg t$, an on--site spin
singlet forms between the f--spin and the conduction electron
spin on each host lattice site. (``Host lattice site'' refers
to an ordinary Kondo site which does not have a Kondo hole.) 
When we put
0, 1, or 2 conduction electrons on the Kondo hole site, the
associated electrons or holes will be localized in the vicinity
of the impurity, and the energies of these three states will be
nearly degenerate. This means that the chemical potential
corresponding to adding a particle or a hole to a half--filled
system is close to zero. This is indeed what we see for 
$J=10$. As $J$ decreases, these midgap states move
toward the edges of the gap as the associated states
become less localized. 

\subsection{Gaps}
We have calculated the spin, charge, and neutral gaps as a function
of $J$ for $L$=8, 16, and 24 at half--filling ($N=L$). 
Our results are shown in Fig.
\ref{fig:gaps}. For comparison we show the corresponding
values of the gaps when there is no Kondo hole. Without
a Kondo hole the spin gap is smaller than the charge gap
for all values of $J$. However, when there is a Kondo hole,
the spin gap is larger than the charge gap for 
$J\stackrel{\textstyle >}{\sim}3$ and smaller than the charge gap for
$J\stackrel{\textstyle <}{\sim}3$. To understand this behavior,
note that when $J\gg t$, we can describe the eigenstates in terms of 
simple on-site states. Each ordinary Kondo site can 
be in a singlet state
involving the f-electron and a conduction electron with an
energy of $-3J/4$, a spin triplet state with energy $J$/4, a
``hole'' state with no conduction electrons 
($S=1/2,I=1/2,I_{z}=-1/2$), or a ``particle''
state with two conduction electrons ($S=1/2,I=1/2,I_{z}=1/2$). 
The particle and hole states have zero energy.
The Kondo hole can have one conduction electron ($S=1/2$, $I=0$),
be in a ``hole'' state with no electrons 
($S=0$, $I=1/2$, $I=-1/2$), or be in a ``particle''
state with 2 conduction electrons 
($S=0$, $I=1/2$, $I_{z}=1/2$). These three Kondo hole states
have zero energy. In the ground state, the
Kondo hole has one conduction electron and every 
site of the host lattice is a singlet when $J\gg t$. 
The lowest spin excitation consists of a single host
site with a spin triplet, with the remaining sites 
being in their ground state configuration; this gives
$\Delta_{S}\approx J$.\cite{ccysrw,zqw,sigrist} The
lowest charge excitation ($S=1/2$, $I=1$)
consists of a hole (particle) on the Kondo 
hole site and a particle (hole) on a host site. Since
one singlet is destroyed, this results in
$\Delta_{C}\approx 3J/4$. Notice that these estimates
indicate that the spin gap is greater than the
charge gap for large $J$. The low-lying eigenstates
consist of linear combinations of these local excitations.
These simple estimates of the
gaps work very well for $J\gg t$; e.g., for $J=100$ and
$L=24$ we find
numerically that $\Delta_{S}=99.9$ and
$\Delta_{N}\cong\Delta_{C}\cong 74.0$, and for $J=10$ and $L=24$ 
we find $\Delta_{S}=9.40$ 
and $\Delta_{N}\cong\Delta_{C}\cong 6.57$.

\subsection{Where the Spin Resides}
At half--filling ($N=L$) a Kondo lattice with a 
single Kondo hole has a total spin $S=1/2$. When
$J\gg t$, the Kondo hole has one conduction electron 
and every site of the host lattice is a singlet in
the ground state. Thus the spin--1/2 resides in the 
conduction orbital of the Kondo hole. This can be seen
in Fig. \ref{fig:spin}a where we plot the z--component
of the conduction spin versus site. As $J$ decreases,
the spin--1/2 is no longer predominantly in the 
conduction orbitals. Rather it is primarily in the
f--orbitals of the sites neighboring the Kondo hole.
For $J=1$, as Fig. \ref{fig:spin}b shows, the 
z--component of the f--spins on the nearest neighbor 
sites are polarized and have most of the spin. 
The f--spins on neighboring sites further away from the Kondo
hole have RKKY oscillations with an envelope that
decays exponentially, indicating that the spin is
localized in the f--orbitals for small $J$. 
Notice that for large $J$, e.g., $J=10$, the spin density
has very little amplitude in the f--spins. We can fit
the absolute value of the z--component $|<0|S_{z}^{f}(r)|0>|$ of 
the f--spins to the form $\exp(-r/\xi)$, where $r$ is the distance
from the Kondo hole and
$\xi$ is the localization length. In 
Fig. \ref{fig:loclength} we plot the localization
length $\xi$ versus $J$. 

This crossover from
conduction spins to f--spins as $J$ decreases is shown
in Fig. \ref{fig:crossover} where we plot the total
conduction spin $\sum_{i}S_{z}^{cond}$ and the
total f--spin $\sum_{i}S_{z}^{f}$ of the lattice
versus $J$. The crossover occurs around $J\approx 4t$.
We can understand why this crossover occurs in the
following way. As we discussed earlier, for large $J$,
the spin is primarily in the conduction orbital of
the Kondo hole. For $J\stackrel{\textstyle <}{\sim}4t$,
it is energetically favorable for both up and down
spin conduction electrons to hop freely. Putting the
spin--1/2 in the conduction spins would polarize the
conduction electrons and impede the
hopping of the down spin electrons. So the spin--1/2
resides primarily in the f--spins. Polarizing the
f--spins costs exchange
energy but that is less important than the kinetic energy
for $J\stackrel{\textstyle <}{\sim}4t$. To see why
the crossover occurs at $J\approx 4t$, note that we can
write the Hamiltonian as 
\begin{equation}
H=-t\sum_{i\sigma}\left(c^{\dagger}_{i\sigma}c_{i+1\sigma}
+h.c.\right)
+\sum_{i}\frac{J(i)}{4}({\vec \sigma}_{if})_{\alpha\beta}
\cdot({\vec \sigma}_{ic})_{\gamma\delta}
f^{\dagger}_{i\alpha}f_{i\beta}c^{\dagger}_{i\gamma}
c_{i\delta}
\end{equation}
where we sum over repeated Greek indices.
The hopping term and the exchange term are comparable when
$J\approx 4t$.

\subsection{Susceptibility}
At zero temperature the uniform susceptibility
$\chi(q=0)$ is zero because there
is a spin gap between the ground state and the lowest spin
excitations. The ground state is an $S=1/2$ doublet whose
energy is linear in a uniform magnetic field due to Zeeman
splitting. Thus $\chi(q=0)=-\partial^{2}E/\partial H^{2}=0$.

However, the staggered susceptibility $\chi(q)$
is finite. To calculate the susceptibility $\chi(q)$, we apply 
a small staggered magnetic
field $h_{z}=h_{o}cos(qr)$ which
couples to both the f--spins and the conduction spins.
The magnitude $h_{o}$ lies between $10^{-6}t$ and $5\cdot 10^{-4}t$.
When $h_{o}$ is this small, the plot
of $S(q)$ versus $h_{o}$ is a straight line whose slope
is the susceptibility $\chi(q)$. ($S(q)$ is the
Fourier transform of $<0|S_{z}(i)|0>$.) We
use periodic boundary conditions with $q_{n}=2\pi n/L$. 


The staggered susceptibility $\chi(q=2k_{F})$
is a measure of the RKKY coupling between f--spins.
($k_{F}$ is the Fermi wavevector
of the noninteracting conduction electrons.)
At half--filling $2k_{F}a=\pi$ and the RKKY coupling favors 
antiferromagnetic alignment of the f--spins. Since the
Kondo hole is missing an f--spin, we expect the RKKY
oscillations, and hence $\chi(q=2k_{F})$, to be
diminished by the presence of a Kondo hole. 
In addition the Kondo hole breaks translational invariance
and allows the system to respond at wavevectors 
other than the wavevectors of the staggered field. 
This will also reduce $\chi(q=2k_{F})$ relative to
its value in lattice without a Kondo hole.

As $J$ approaches zero, numerical noise can induce spurious
antiferromagnetic ordering of the lattice because the
states with and without long range order are very close
in energy. Because of this, we only study short chains
(4 and 6 sites) which we can diagonalize exactly.
In Fig. \ref{fig:chivsJ} we plot the staggered susceptibility
$\chi(qa=2k_{F}a=\pi)$ versus $J$ for lattices with and without
a Kondo hole at half--filling.\cite{stagger} As expected, 
our results show that the
staggered susceptibility is greatly reduced by the presence of
a Kondo hole. This reflects the suppression of RKKY oscillations.
As the lattice gets longer, we expect that the effect of a single
Kondo hole will be diluted and $\chi(q)$ will approach the value
of a pristine lattice. 

\section{Conclusions}
We have studied a Kondo hole in the middle of
a one--dimensional 
Kondo lattice at half filling using the density 
matrix renormalization group technique. The Kondo
hole introduces midgap states which move from the
middle of the quasiparticle gap to
the edges as the exchange coupling $J$ goes from
large values to small values. As $J\rightarrow \infty$,
the chemical potential of
these midgap states goes to zero which corresponds 
to the degeneracy of
states with 0, 1, and 2 conduction electrons on the
Kondo hole site. At half filling there are an odd number
of spins and the ground state has $S=1/2$. This 
spin--1/2 is localized in the vicinity of the Kondo
hole. It is primarily in the f--spins for small
$J$ and in the conduction spins for large $J$.
The crossover occurs at $J\approx 4t$ where the
kinetic energy is comparable to the exchange energy.
We presented results on the spin gap, charge gap,
and neutral gap as a function of $J$. For small $J$
the spin gap is smaller than the charge gap. However, 
for large $J$, the spin gap is larger than the charge
gap because the energy to change a singlet into a 
triplet is $\Delta_{S}\sim J$ while the energy
to transfer an electron or a hole from the Kondo hole
to another site which had a singlet is $\Delta_{C}\sim 3J/4$. 
The presence of the Kondo hole reduces RKKY 
oscillations as can be seen in the staggered susceptibility
$\chi(qa=2k_{F}=\pi)$.

Putting Kondo holes into Kondo insulators can be done
experimentally by replacing Ce ions with La ions. It should 
be possible to look in real materials for some of 
the effects listed above, e.g., midgap states and reduced
RKKY oscillations. However, having a finite concentration of
Kondo holes introduces effects that we have not considered
here such as impurity bands and interactions between 
the Kondo holes.\cite{impbands}

\section*{Acknowledgements}
We would like to thank Herv\'{e} Carruzzo and Jon Lawrence for
helpful discussions.
This work was supported in part by ONR grant  
N000014-91-J-1502 and
an allocation of computer time from the 
University of California, Irvine.

\newpage
\begin{figure}
\caption{Chemical potential scaled by $J$ versus number of
conduction electrons $N$. The scaled chemical potential
is defined by $\mu=[E(N)-E(N-1)]/J$. $t=1$, $L=16$ and 
the Kondo hole is on site $i=8$. Open boundary conditions
are used. The midgap states
associated with large $J$ move toward the edges of the
gap as $J$ decreases. For comparison, we show the
chemical potential for the case of no Kondo hole with
$J=10$. The solid lines are guides to the eye.}
\label{fig:chempot}
\vspace*{0.9cm}

\caption{(a) Spin and charge gaps versus $J$ 
for $L=8$, 16, and 24 sites with open boundary
conditions.
The Kondo hole is on site $i=L/2$. $t=1$. For
comparison, we show the spin and charge gaps 
without a Kondo hole for $L=24$. 
(b) Neutral gap versus $J$ for $L=8$, 16, and 24 sites
with open boundary conditions.
The Kondo hole is on site $i=L/2$. $t=1$. The solid
lines are guides to the eye.}
\label{fig:gaps}
\vspace*{0.9cm}

\caption{(a) z--component of the conduction spin
versus site for $J=1$ and $J=10$ for a 24 site 
lattice with the Kondo hole on site $i=12$.
(b) z--component of the f--spin
versus site for $J=1$ and $J=10$ for a 24 site 
lattice with the Kondo hole on site $i=12$.
Notice that the amplitude of the RKKY oscillations 
fall off exponentially for $J=1$. The solid lines are
guides to the eye.}
\label{fig:spin}
\vspace*{0.9cm}

\caption{Total f--spin ($S_{z}^{f}=\sum_{i}S_{z}^{f}(i)$), 
total conduction spin ($S_{z}^{cond}=\sum_{i}S_{z}^{cond}(i)$), 
and the difference ($S_{z}^{f}-S_{z}^{cond}$) versus $J/t$.
Notice that where the spin--1/2 resides crosses over
from f--spins to conduction spins as $J$ increases. The solid
lines are guides to the eye.}
\label{fig:crossover}
\vspace*{0.9cm}

\caption{Spin localization length $\xi$ versus $J$ for a
24 site lattice with open boundary conditions. $\xi$ is
deduced by fitting $<0|S_{z}^{f}(r)|0>$ to the form
$\exp(-r/\xi)$ where $r$ is the distance from the hole.
The error bars are the standard deviation of the fit. 
The error bars are smaller than the size of the points
for all $J$ except $J=0.75$.
The solid line is a guide to the eye. $t=1$.}
\label{fig:loclength} 
\vspace*{0.9cm}

\caption{Staggered susceptibility $\chi(qa=2k_{F}a=\pi)$
versus $J$ for $L=4$ and 6 at half--filling with periodic
boundary conditions. The Kondo hole is on site $i=L/2$.
For comparison we also show $\chi(qa=2k_{F}a=\pi)$ for
lattices without a Kondo hole.}
\label{fig:chivsJ}
\vspace*{0.9cm}

\end{figure}


\begin{thebibliography}{99}
\bibitem{neutral} Because the spin is 1/2 and not zero,
we do not call the particle--hole excitation
($S=1/2$, $I=0$) a neutral singlet, but rather just a neutral
excitation.
\bibitem{stagger} When we applied a staggered field ($q=\pi$) to
a 4 site lattice with a Kondo hole on site $i=2$, the field pointed
up at the Kondo hole and down on the nearest neighbor sites. Since
the net spin $S=1/2$ is in the conduction orbital on the Kondo hole
for large $J$, the ground state has $S_{z}=+1/2$ for $J\geq 2.7$. 
For small $J$ the net spin resides primarily in the f-spins
near the Kondo hole. To align with the staggered field, the
f-spins on the nearest neighbors to the Kondo hole point down.
So the ground state has $S_{z}=-1/2$ for $J\leq 2.6$.
A similar crossover in the z--component of the spin of the 
ground state occurs for $L=6$.
\end{thebibliography}
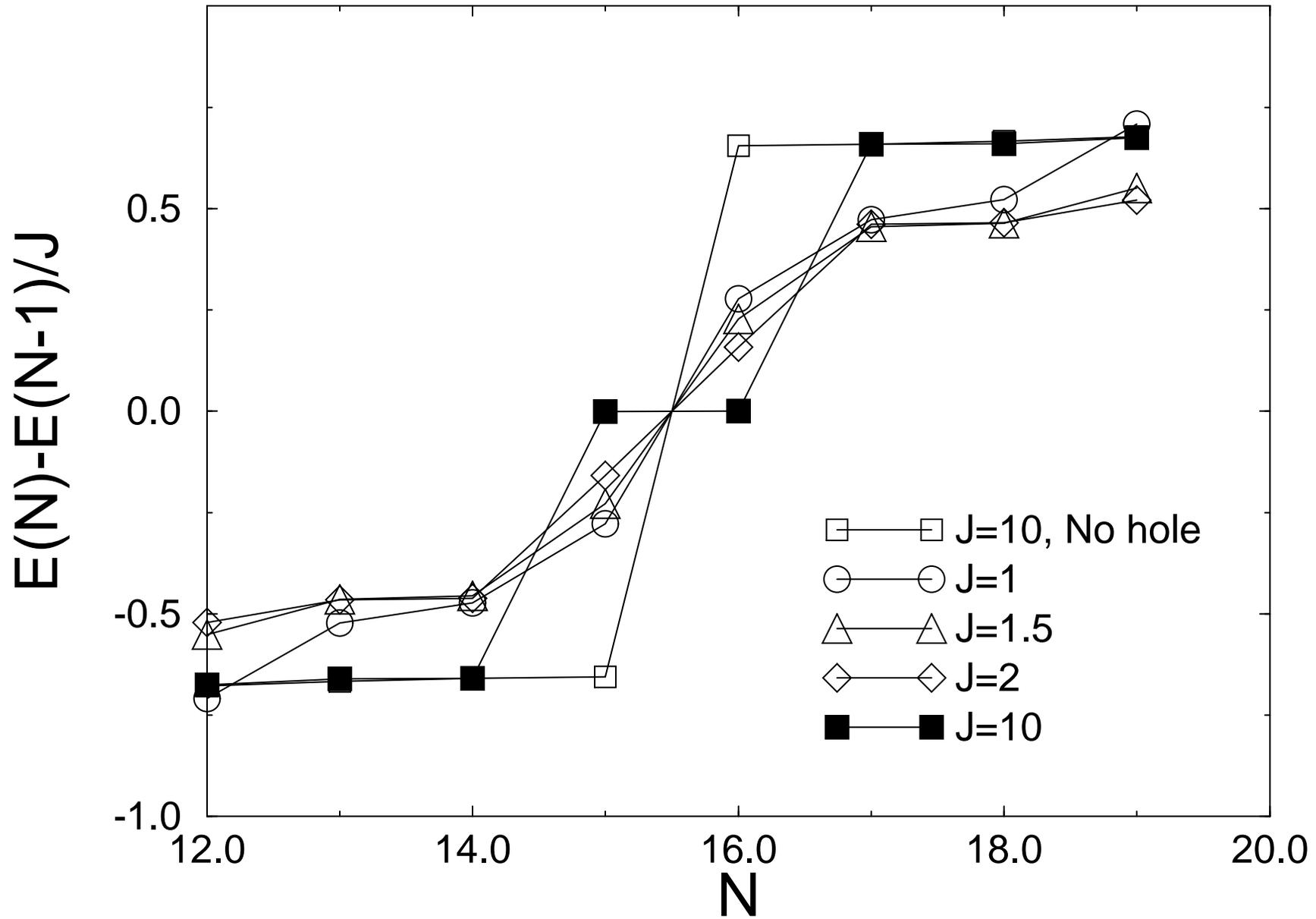
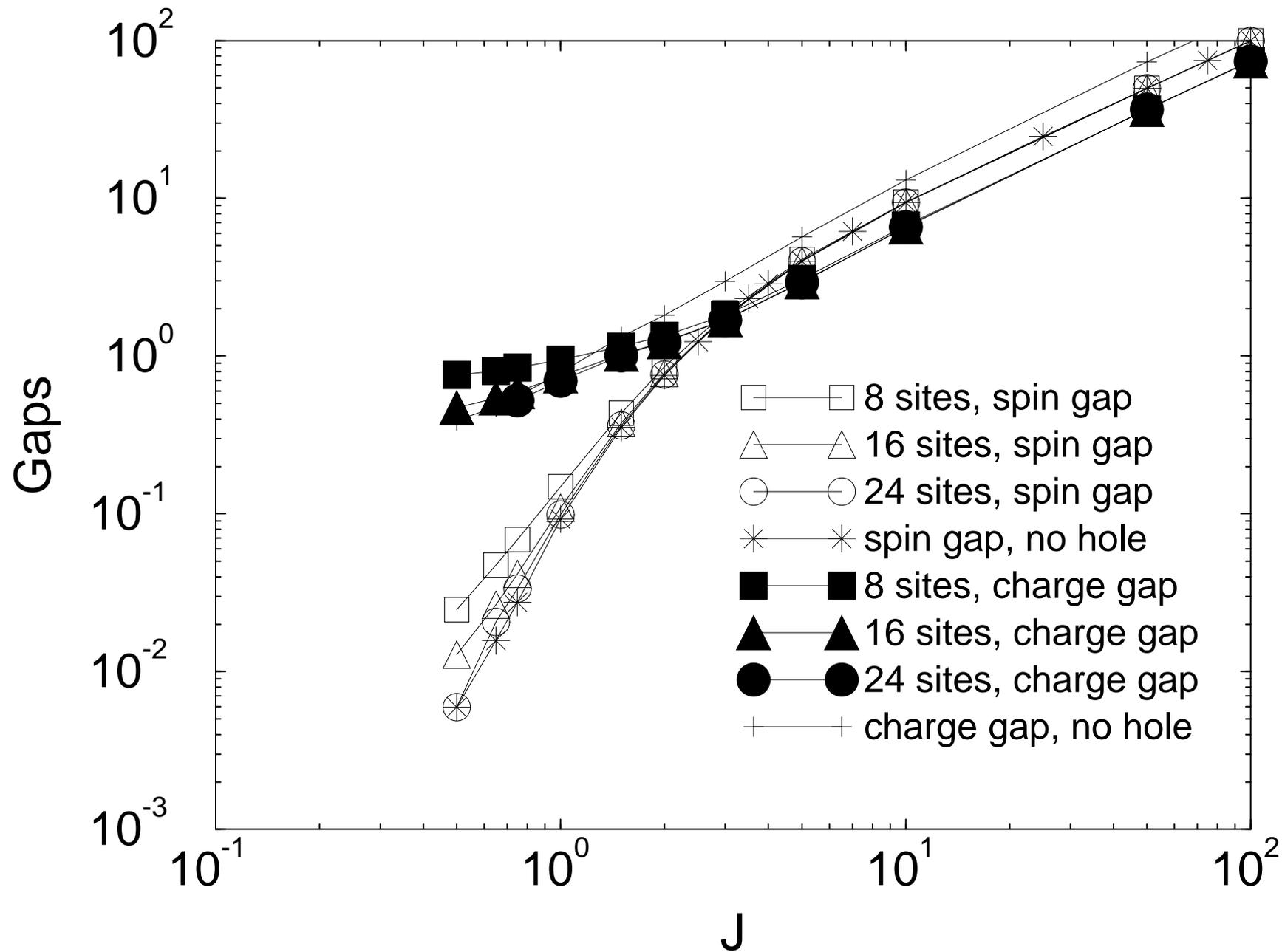
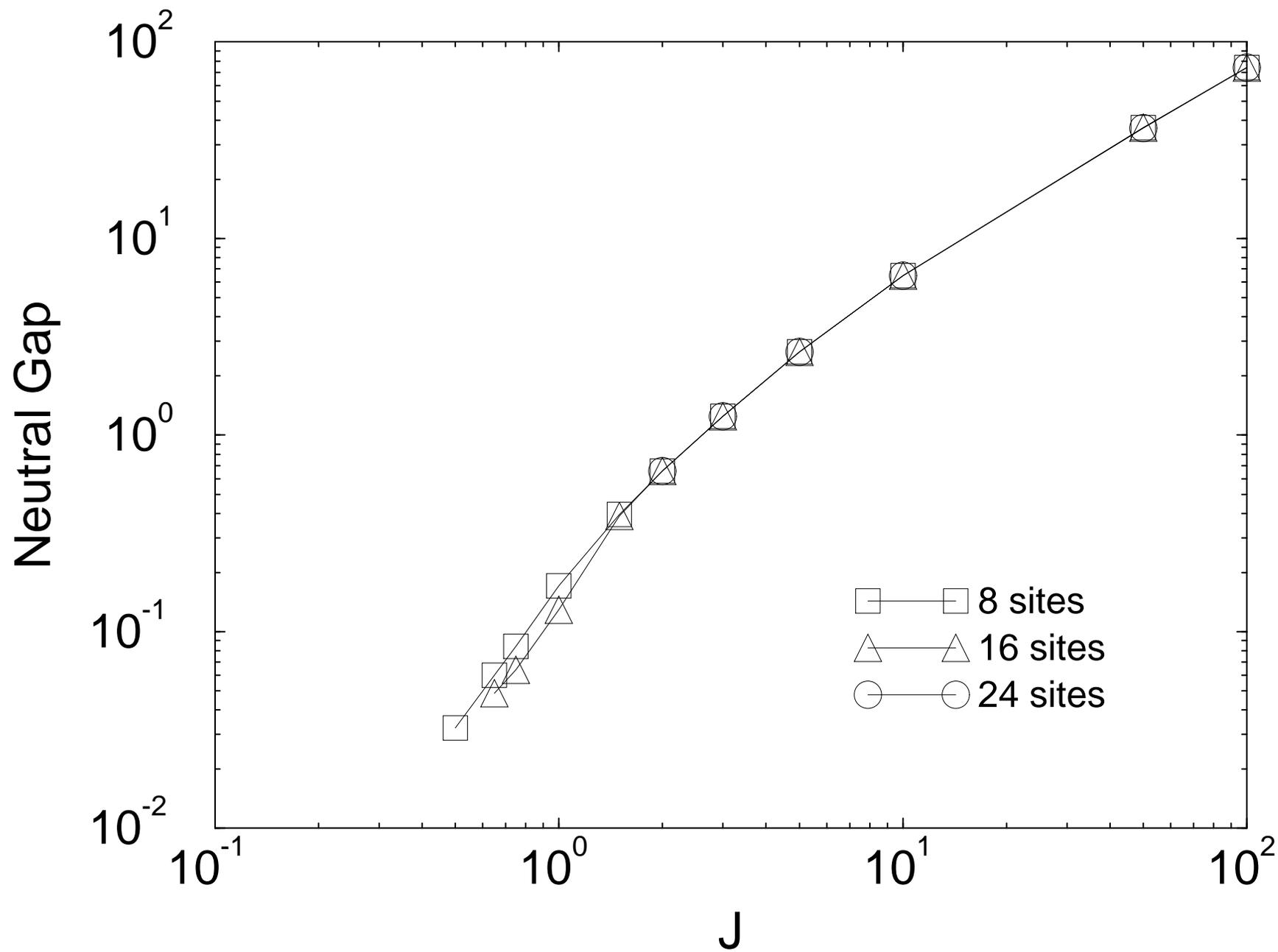
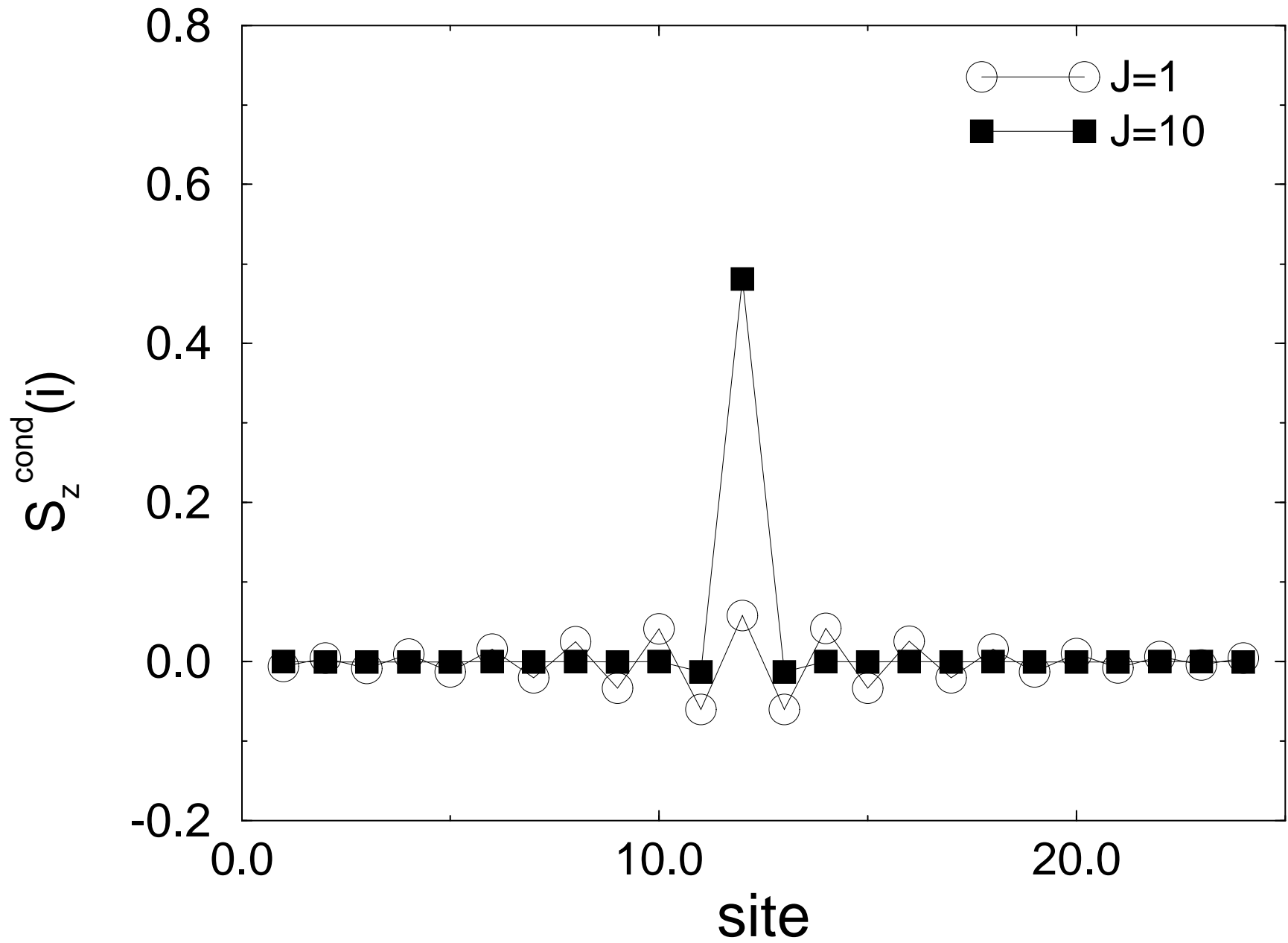
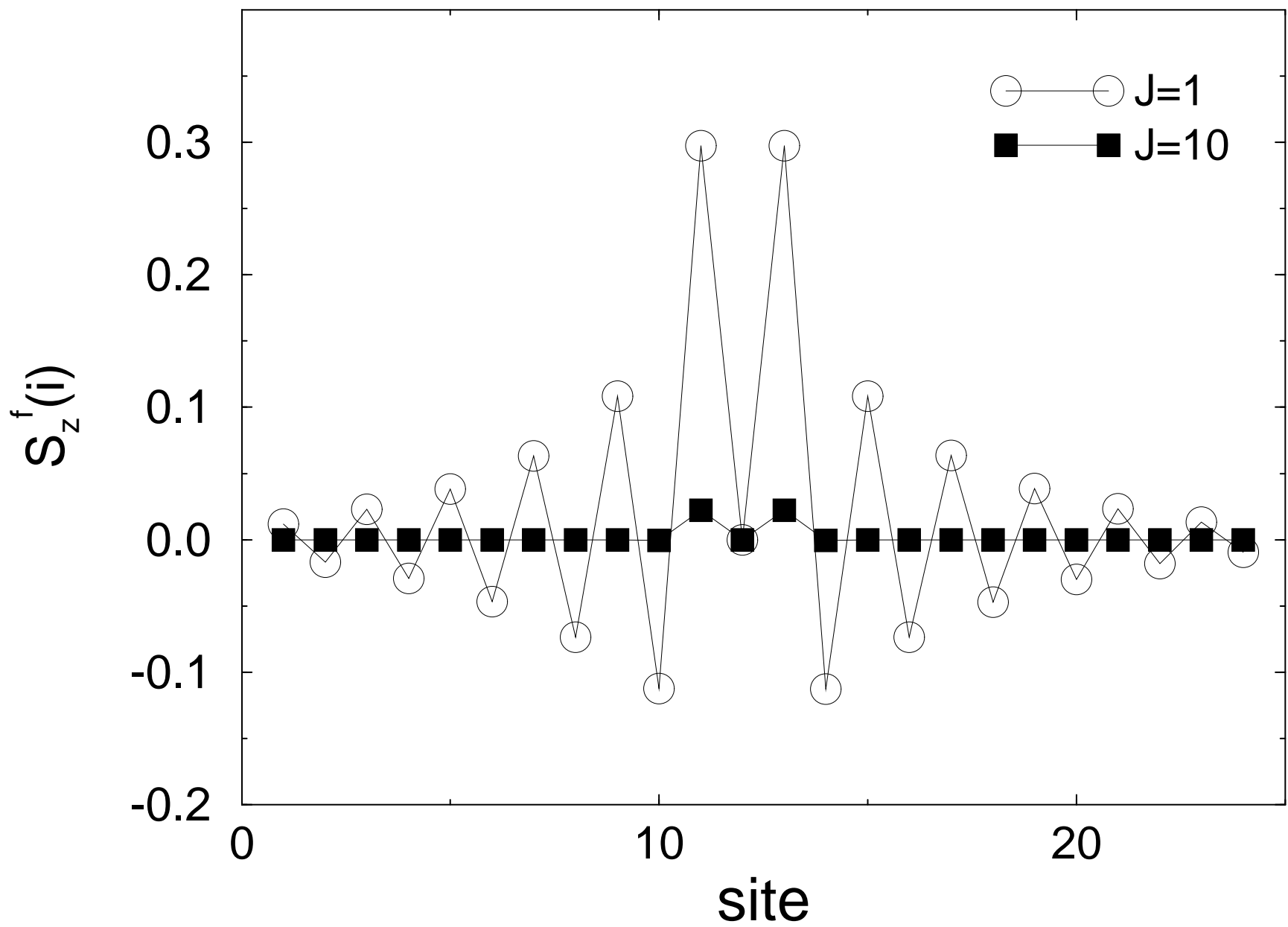
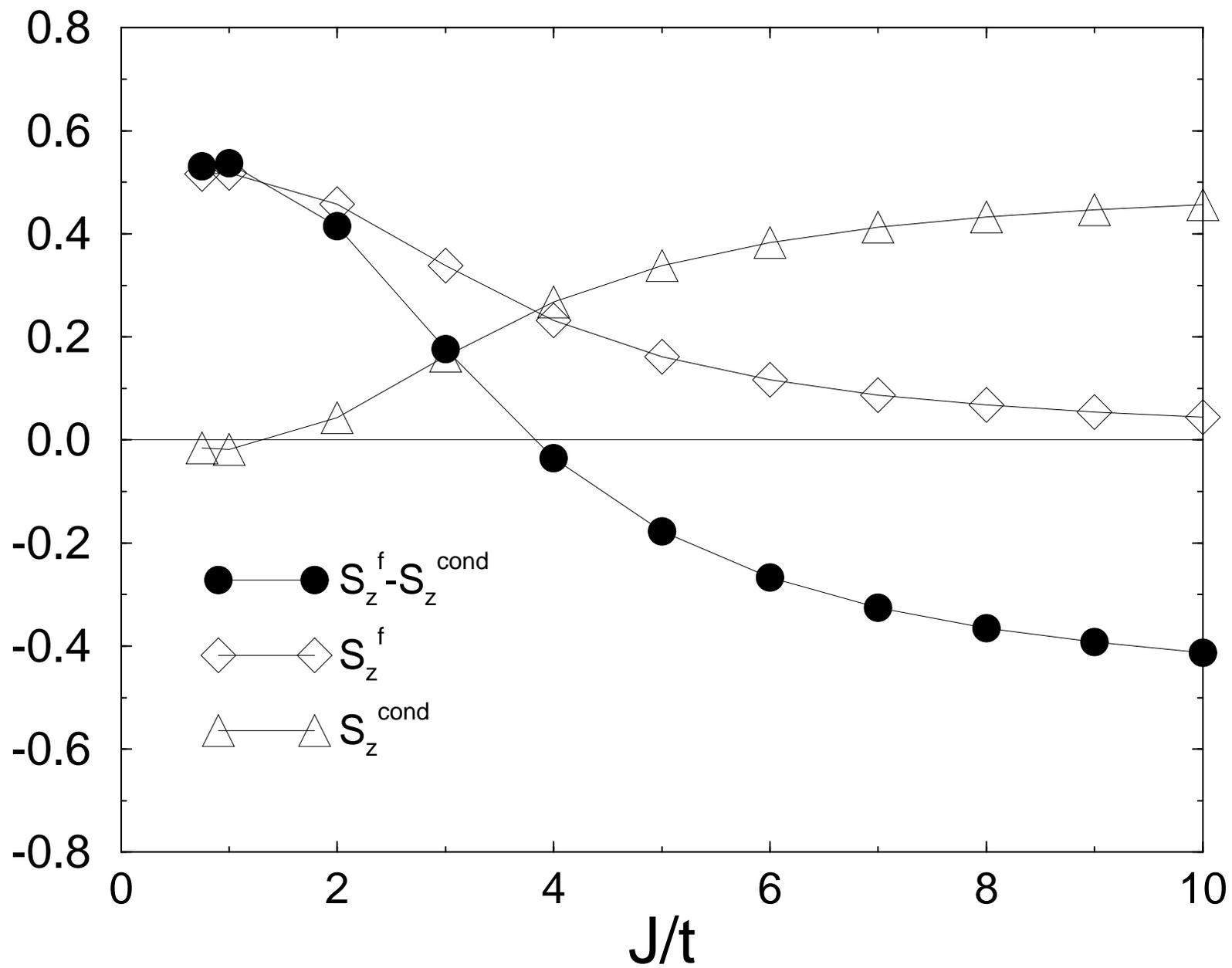
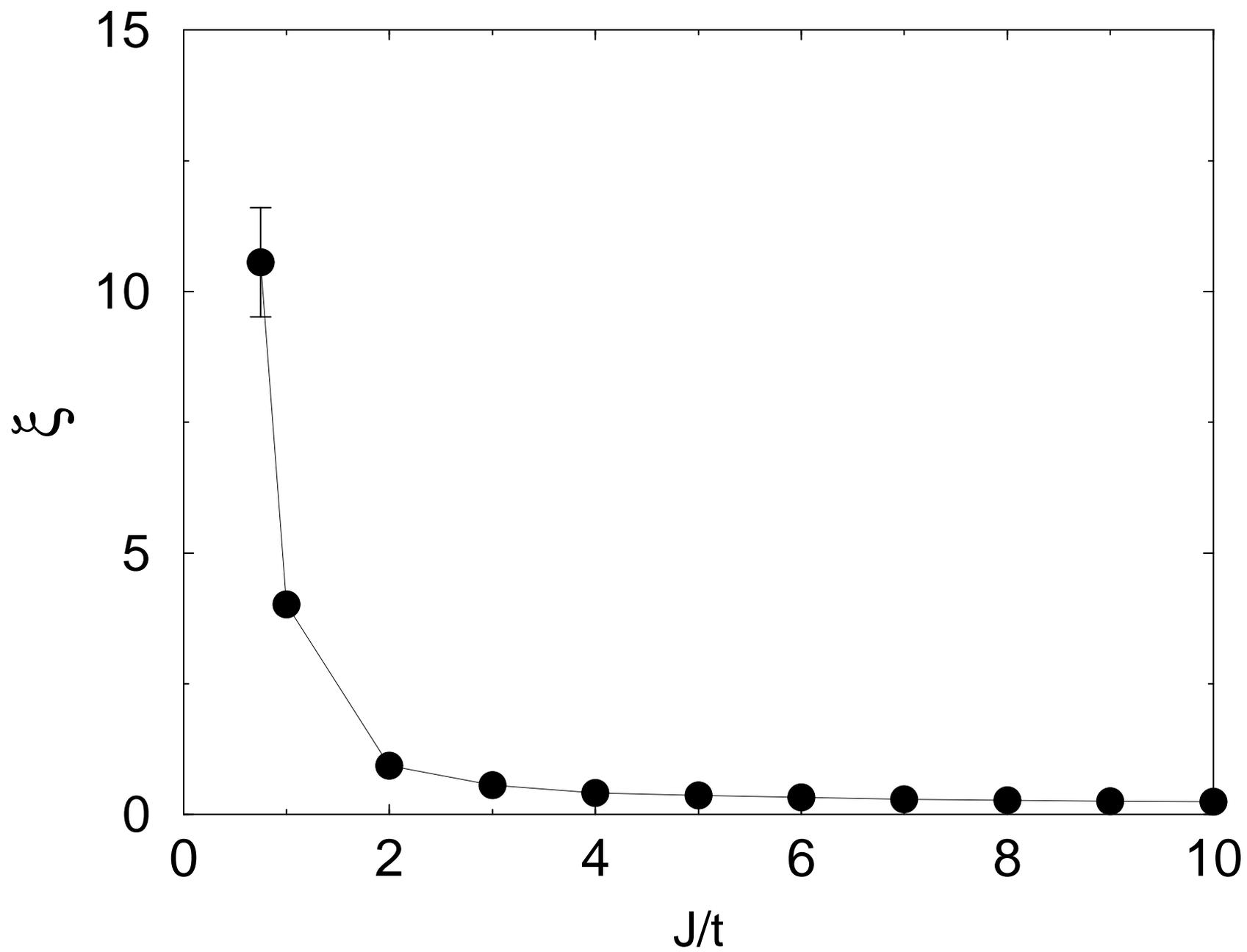
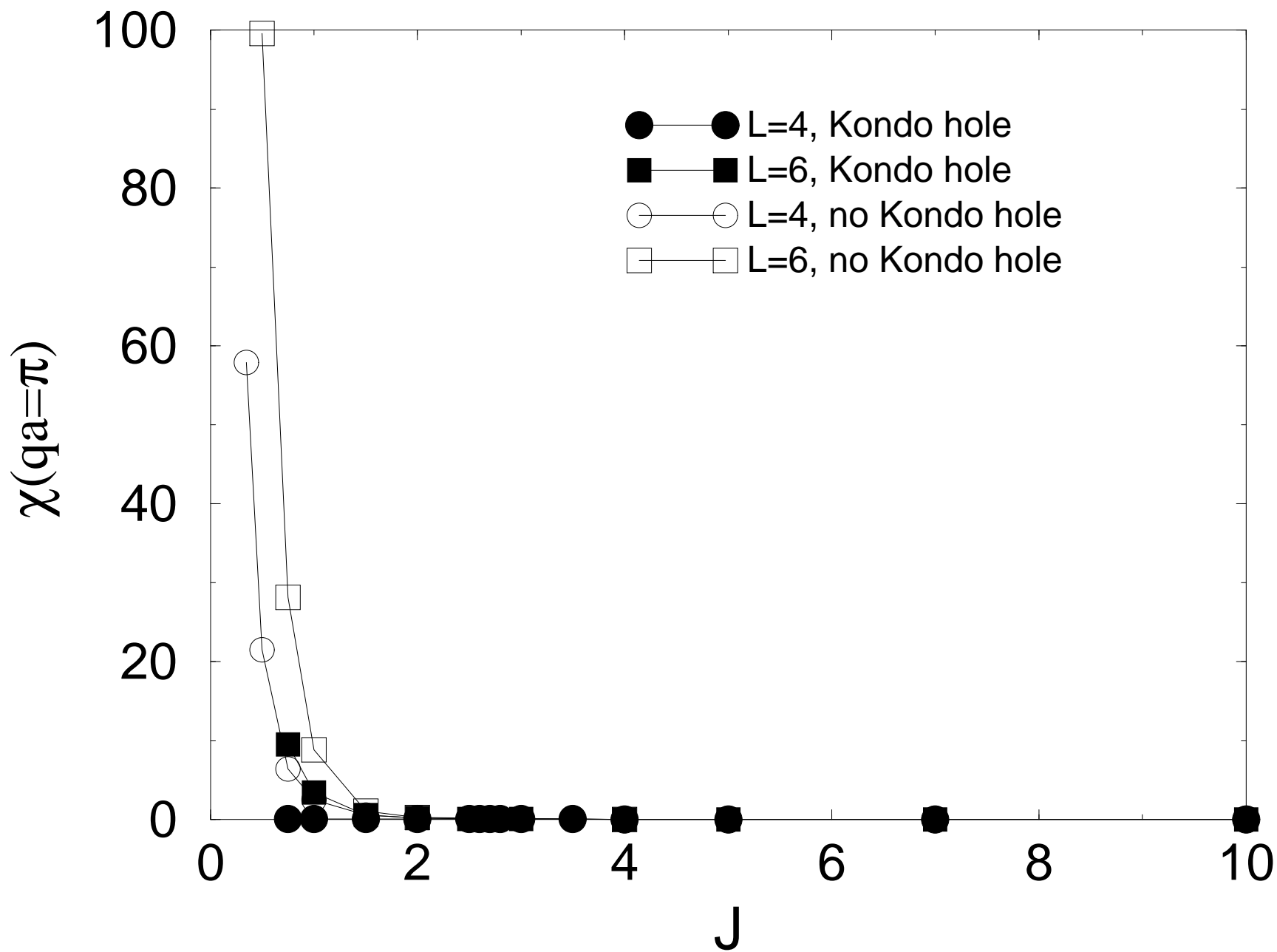
\end{document}